\documentclass[bibyear]{aa}

\usepackage{graphicx}
\usepackage{txfonts}
\usepackage{natbib}

\begin{document}
\title{A search for magnetic fields on central stars in planetary nebulae}
\author{F. Leone\inst{1}, R.L.M. Corradi\inst{2,3},  M.J. Mart\'{i}nez Gonz\'alez\inst{2,3}, A. Asensio Ramos\inst{2,3}, \and R. Manso Sainz\inst{2,3}
}
\institute{
Universit\`a di Catania, Dipartimento di Fisica e Astronomia, Sezione Astrofisica, Via S. Sofia 78, I--95123 Catania, Italy
\and
Instituto de Astrof\'{\i}sica de Canarias, 38205, La Laguna, Tenerife, Spain
\and
Departamento de Astrof\'{\i}sica, Universidad de La Laguna, E-38205 La Laguna, Tenerife, Spain
}
\authorrunning{Leone et al.}

\date{Received To be inserted later, accepted To be inserted later}

\abstract
{One of the possible mechanisms responsible for the panoply of shapes
in planetary nebulae is the presence of magnetic fields that drive the
ejection of ionized material during the proto-planetary nebula
phase.}
{Therefore, detecting magnetic fields in such objects is of key
importance for understanding their dynamics. Still, magnetic fields
have not been detected using polarimetry in the central stars of
planetary nebulae.}
{Circularly polarized light spectra have been obtained with the {\it Focal Reducer and Low Dispersion Spectrograph}
at the Very Large Telescope of the European Southern Observatory and the
{\it Intermediate dispersion Spectrograph and Imaging System} at the William Herschel Telescope.
Nineteen planetary nebulae spanning very different morphology 
and evolutionary stages have been selected. Most of central stars have been observed
at different rotation phases to point out evidence of magnetic variability.}
{ In this paper, we present the result of two observational campaigns
aimed to detect and measure the magnetic field in the central stars of
planetary nebulae on the basis of low resolution spectropolarimetry.
In the limit of the adopted method, we can state that large scale
fields of kG order are not hosted on the central star of  planetary nebulae.}
{}


\maketitle

\section{Introduction}

In the last twenty years, extensive ground-based and HST imaging has
revealed the extraordinary shapes of planetary nebulae (PNe).  The
original paradigm of PNe as spherical shells expanding uniformly
around the stellar remnant, a hot white dwarf, is clearly far from
describing reality. We now know that different kind of geometries,
from spherical or ellipsoidal to highly collimated ones, characterize
the overall shape of PNe \cite[e.g.][]{Corradi95}.  In addition, on smaller scales -- embedded in
the main bodies or external to them -- a rich realm of additional
structures is found, such as symmetrical (or not) pairs of knots,
jets, ansae, etc.  \citep{Gon01}.

These articulated morphologies clearly indicate that mass loss at the
very end of the asymptotic giant branch (AGB) is complex and still far
to be fully understood. Presently, a strong dynamical interaction
between the massive/slow AGB wind (produced by surface levitation of
gas due to convection, stellar pulsations and radiation pressure on
dust) and the fast/tenuous post-AGB wind (driven by blanketed UV
absorption lines of ions) is though to play a basic role in governing
the formation and evolution of PNe, but it does not explain the
deviation from the spherical geometry. In order to account for it, a
number of explanations have been proposed \cite[see e.g.][]{Balick_Frank2002},
the most popular ones being interactions in
mass-exchanging binary systems or the emergence of surface magnetic
fields. Such models have one preferred symmetry axis and, hence, might
account for simple bipolar flows, but in order to reproduce all
multi-polar structures (like multi-lobal nebulae or multiple blobs and
jets) the theoretical scenario had to be enriched with other
ingredients able to break the axial symmetry (like precession of
accretion disk winds or off-axis magnetic fields).

The binary scenario has been explored both theoretically
(e.g. \citealt{Soker01}) and observationally
(e.g. \citealt{Miszalski12}, \citealt{Boffin2012}), and while it can
explain a number of structures observed in the nebulae, it is unlike
that gravitational interactions alone are sufficient to explain all
the various outflows that are sometimes found in the same nebula
(cf. e.g. the case of Mz3 in Santander et al. \cite{Santander04},
where four distinct flows with increasing collimation degree have been
identified).

Magnetic fields can provide the additional mechanisms needed to
explain the observed structures. With this aim, a variety of
magneto-hydrodynamical (MHD) simulations of the nebular shaping have
appeared in the literature (e.g. Garcia-Segura et
al. \citealt{Garcia99}; Matt et al. \citealt{Matt04}; Frank \&
Blackman \citealt{Frank04}). The magnetic field may be either a fossil
remnant from the progenitor on the main sequence (e.g. Ap stars), or
can be generated by a dynamo at the interface between a rapidly
rotating stellar core and a more slowly rotating
envelope. \cite{Blackman01} argue that some remnant field anchored in
the core will survive even without a convection zone, although the
convective envelope may not be removed completely. Thomas et
al. \cite{Thomas95} have shown that white dwarfs which do have thin
surface convection zones can support a near-surface dynamo. Since the
field strength in their model is higher at higher luminosities, this
would particularly be true for central stars of PNe.  That some
central stars should contain significant magnetic fields is also
indicated by the fact that some 10-30$\%$ of all white dwarfs have
magnetic fields between 10$^3$ and 10$^9$ Gauss.

In spite of these facts, and that the MHD simulations are quite
successful in reproducing several of the observed nebular structures,
to date very little observational evidence has been obtained of the
existence of such magnetic fields.   {\bf Still, magnetic fields
have not been detected using in the central stars of planetary nebulae.} First positive detections at a
kiloGauss level were claimed by Jordan et al. \cite{Jordan05} in two
PNe, but Leone et al. \cite{Leone2011} first, and then Bagnulo et al. \cite{Bagnulo12}
and Jordan et al. \cite{Jordan12} could not confirm these results.

Jordan et al. \cite{Jordan12} have measured the magnetic field in the central stars of eleven
planetary nebulae concluding that to date there is still no evidence
for the existence of magnetic fields in PN central stars.
In this paper, we continue our search for magnetic fields in PN
central stars via spectropolarimetry, as done in Leone et
al. Leone et al. \cite{Leone2011}. The survey is not focussed on specific morphological
classes, but should be considered as a panoramic view with the main
goal of highlighting the overall properties of magnetic fields (if
any) in the central stars of PNe in general.  Detecting magnetic
fields in one or the other of the observed morphologies would
demonstrate that magnetic fields are indeed at work, and would start
revealing which kind of shaping processes they are relevant for.
With the aim to improve the variety and for an omogeneousy
reduction and analysis, spectropolarimetric data for seven
central stars with the FOcal Reducer and low dispersion Spectrograph (FORS)
{\bf have been obtained from the archive of the European Southern Observatory.}

\section{Observations, data reduction and magnetic field measurements}

Measuring stellar magnetic fields is one of the most demanding
techniques because of the need to reach very high signal to noise
ratios. In the weak-field approximation for stellar atmospheres
(Landstreet \citealt{Landstreet1982}; Mathys \citealt{Mathys1989}), the disk-integrated Stokes-V parameter
(the difference between the opposite circular polarized intensities) across spectral line profiles is proportional
to the longitudinal component of the magnetic field integrated over the stellar disk, the so called {\it effective magnetic field}.
{\bf High-resolution circular spectropolarimetry gives the possibility to distinguish photospheric regions with
positive and negative magnetic fields (Leone \& Catanzaro \citealt{Leone2004}; R=115,000).
Circular spectropolarimetry is also useful at moderate resolution (Leone \& Catanzaro \citealt{Leone2001}; R = 15,000)
to detect magnetic fields,  but it is still prohibitive for faint stars.} As to white dwarfs, Angel \& Landstreet \cite{Angel1970}
introduced a method based on narrowband ($\sim$30 Å) circular
photopolarimetry on the wings of the H$_{\gamma}$ Balmer line. 
Bagnulo et al. \cite{Bagnulo2002} have shown how to co-add the Stokes-V
signal from spectral lines, as observed at low resolution, and measure the effective field of spectra lines
on very faint stars.

With the aim to measure the effective magnetic field of the central stars of planetary nebulae,
we have adopted the procedures and methods presented in Leone et al. \cite{Leone2011}.
Because of the necessary huge signal to noise ratio (S/N $\sim$5000 was achived for an upper limit
of 300~G in the case of PNe NGC1360),  we have selected PNe among the brightest ones,
with the additional criterion of covering a range of morphologies as
large as possible. Another important aspect to detect magnetic fields
with the adopted technique is to select targets for which the nebular
Balmer emission is negligible long the line of sight of the central
star.

Spectropolarimetric data have been collected 1) at the William
Herschel Telescope (WHT) at the Observatorio del Roque de los
Muchachos at La Palma, Spain, using the ISIS spectrograph and 2) at
the Unit 1 of the Very Large Telescope at ESO, Chile, using FORS2.
With ISIS, data were obtained in the 3785 - 4480 \AA\, range at
resolution R = 5000 with the procedures described in Leone \cite{Leone2007}.
As to FORS2, data were obtained in the 3800 - 5200 \AA\, range at
resolution R = 2700 with the procedures described in Leone et al. \cite{Leone2011}.

Information about the basic properties of each target PN, a log-book of the observations and obtained
results are listed in Table 1. 

\begin{figure*}
   \centering

\includegraphics[width=\textwidth,height=18cm]{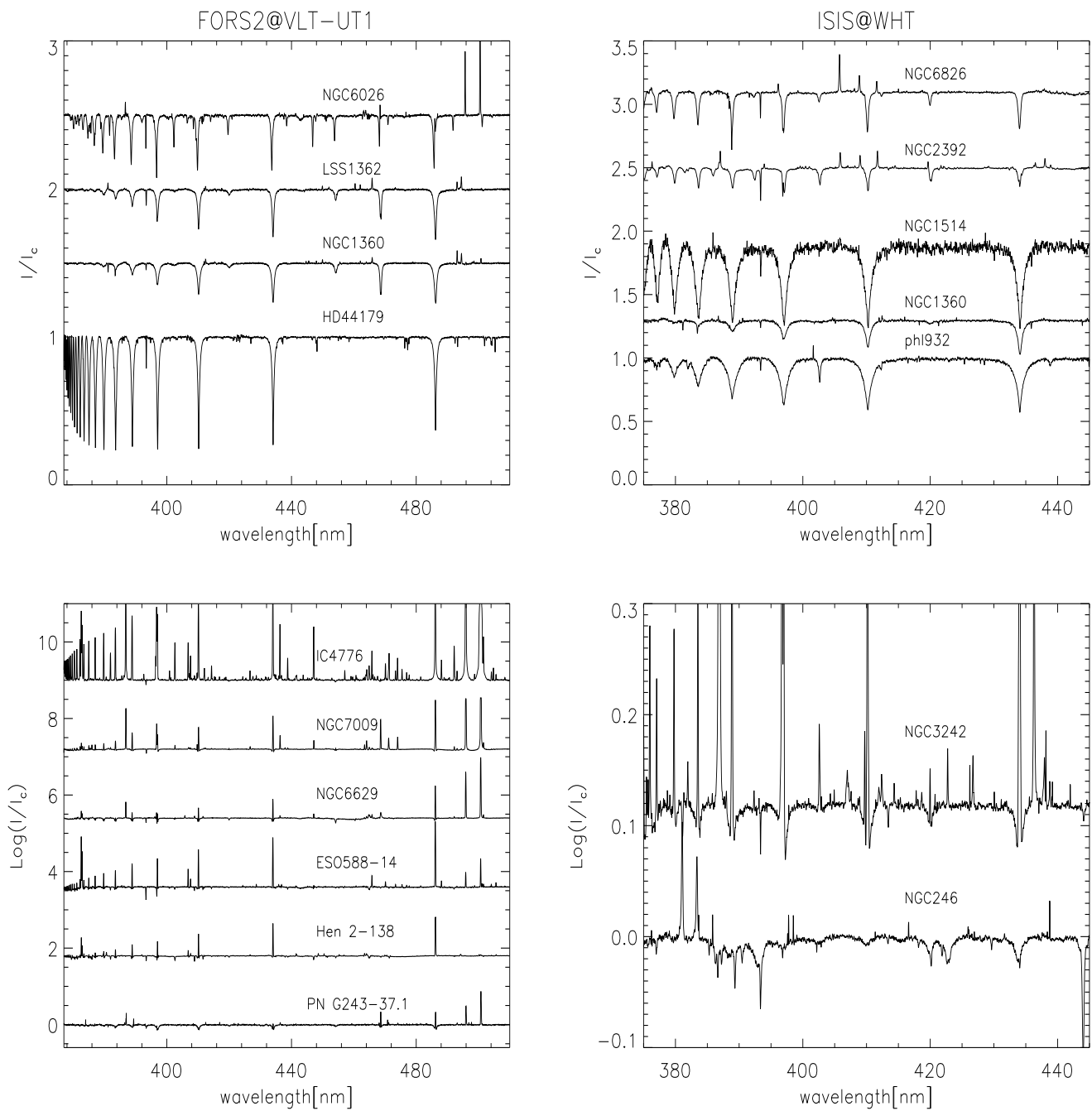}
\caption{Observed central stars of planetary nebulae listed in Table~1.
Left panel shows the stars observed with FORS2@UT2.
Right panels show the stars observed with ISIS@WHT.
Low panels show the logarithimic relative intensity for those stars
dominated by emission lines. An ad hoc shift has been assumed to avoid the overlapping of spectra.}
\end{figure*}

In order to increase the signal to noise level of the polarized
spectra, we averaged the observations of the same object that are not
separated more than half the period of rotation of the object. Each
averaged observation has been labeled with a number from 1 to 10.  The
reduction and the demodulation procedure are the same for all objects
(see Leone et al. \citealt{Leone2011}).  Figure 1 shows
the Stokes I spectra as observed with FORS2 and WHT. Last column of Table~1 shows the results of our
spectropolarimetric analysis. No magnetic field has been detected on the central stars
of the selected planetary nebulae. It is worthy to note that for nine targets no analysis could be done either
because the nebular contamination was too high: NGC246, PN G243-37.1, NGC 3242, He2-138, Hen 2-194, ESO588-14, NGC~6629,
IC 4776 and  NGC~7009, lower panels of Fig.\,1.

\begin{table*}
\caption{Main properties of the target PNe.  The B$_{\rm mag.}$ of the central star is reported.
With the present method, the magnetic field could not be measured in stars with a nebular emission overimposed to Balmer lines.}
\label{tab:nebulinfo}
\centering
\scriptsize
\begin{tabular}{lclccr} 
Target    &  B$_\mathrm{mag}$  &  & \multicolumn{3}{c}{Nebular morphology / Comments } \\
 ~~~~~RA ~~(2000)~~ DEC    &                    & Tel.   &     JD-2450000.           & $\sigma \times 10^{-3}$ & \multicolumn{1}{c}{Field [G] } \\
\hline
\hline
NGC 246 = HIP 3678 & 10.2 &  & \multicolumn{3}{c}{ elliptical - \bf emission lines }   \\
00 47 03.3\hspace{0.2cm} $-$11 52 18.9  &       &  WHT & 3718.349             &  &  non measurable\\
\hline
PHL 932 = HIP 4666  & 12.0  & & \multicolumn{3}{c}{asymmetric -  not PN? (see text)} \\
00 59 56.7\hspace{0.2cm}  $+$15 44 13.8    & & WHT &  4011.503  &    1.8 &  $310 \pm  1150$ \\
\hline
NGC 1360 = CPD-26 389     & 11.0 && \multicolumn{3}{c}{prolate ellipsoid + jets (i=$60^\circ$)}  \\
03 33 14.6\hspace{0.2cm}  $-$25 52 18.0 &  & WHT & 4011.647 &  1.0  & $  356\pm  1000 $ \\
 &  &     & 4012.625 &  0.9  & $ -105\pm 1045 $ \\
 &  & VLT & 5553.558 &  0.1  & $  155\pm  115 $ \\
\hline
NGC 1514 = HD 281679      & 10.0 & & \multicolumn{3}{c}{distorted ellipsoid + shell (i=$59^\circ$); binary?; A-type spectrum}\\
  04 09 16.7\hspace{0.2cm}   $+$30 46 28.8    &      & & \multicolumn{3}{c}{two offset rings at 24$\mu$ (WISE)} \\
                          &      & WHT &3718.425 / 3719.489             & 0.3  &  $120\pm 70$ \\
                          &      & &4010.718 / 4012.735 / 4013.598 & 2.4  & $ -680\pm 300 $ \\
\hline
PN G243-37.1             & 11.4 & &   \multicolumn{3}{c}{irregular rim + shell - \bf emission lines } \\
05 03 01.73\hspace{0.2cm}  $-$39 45 44.5          &                    &  VLT &  6174.835        &  &  non measurable\\
\hline
HD 44179 = Red Rectangle  & 9.3  & &\multicolumn{3}{c}{bipolar (reflection) (edge-on)  proto-PN; binary P=322 days}\\ 
06 19 58.2\hspace{0.2cm} $-$10 38 14.7                          &      & &\multicolumn{3}{c}{A1 post-AGB star + low L WD} \\
                          &      & VLT &6175.824   & 0.1 &  $  70  \pm  35 $\\
\hline
NGC 2392 = Eskimo nebula & 10.3 & &\multicolumn{3}{c}{distorted elliptical rim + shell  + knotty disk + caps + jets (pole-on)}  \\
07 29 10.8\hspace{0.2cm}  $+$20 54 42.5       &      &  WHT        & 3718.562/3719.636            &  0.5 & $360\pm500$ \\
\hline
LSS 1362 = TYC 8181-2925-1 & 12.5  & &  \multicolumn{3}{c}{elliptical amorphous; evolved PN}\\
09 52 44.5\hspace{0.2cm}   $-$46 16 51  & &  VLT   & 5553.832  &  0.5 & $-335\pm285$\\
\hline
NGC 3242 = HD 90255 & 10.3  &   &  \multicolumn{3}{c}{ elliptical rim + shell + ansae + halo -  \bf emission lines} \\
10 24 46.1\hspace{0.2cm} $-$18 38 32.6 &   & WHT  & 3718.731  &  &  non measurable \\
\hline
NGC 4361 = HD 107969       & 12.8  & & \multicolumn{3}{c}{mildly bipolar} \\
12 24 30.8\hspace{0.2cm}  $-$18 47 06.4 &    & VLT$^{EJ}$ & 3526.577 &  3.9 & $  655\pm 395$\\
\hline
Abell 36 = HIP 66732       & 11.3  & & \multicolumn{3}{c}{large elliptical with ISM interaction; evolved PN}\\
13 40 41.4\hspace{0.2cm}  $-$19 52 54.9  & &  VLT$^{EJ}$ & 3525.496 &   7.1   &      $ -50\pm     290$\\
                                        &  &  VLT$^{EJ}$ & 3536.472 &   5.8  & $ -240\pm 270$\\
\hline
LSE 125  = PN G335.5+12.4  & 12    & &  \multicolumn{3}{c}{round} \\
15 43 05.4\hspace{0.2cm} $-$39 18 14.6  &   & VLT$^{EJ}$ & 3525.736 &   0.3 & $   10\pm 150 $\\
                                        &  & VLT$^{EJ}$ &  3527.776 &   4   & $ -845\pm 395$\\
\hline
Hen 2-138 = HD 141969  & 10.9  & &  \multicolumn{3}{c}{elliptical with "wiggles"; young - \bf emission lines}\\
 15 56 01.7\hspace{0.2cm  $-$66 09 09.2}     &       &  VLT &    6176.549           &  &  non measurable\\
\hline
NGC 6026 = PN G341.6+13.7                    & 13.2   & &  \multicolumn{3}{c}{amorphous + double-degenerate P=0.53 days (i=$80^\circ$)}  \\
16 01 21.1\hspace{0.2cm} $-$34 32 35.8 &        & VLT & 6173.515/6174.545/6175.504 &  2& $ -220  \pm  130 $ \\
\hline
Hen 2-194 = PN G350.9+04.4 & 13.7 &  & \multicolumn{3}{c}{round rim + shell + bubble - /bf emission lines} \\
17 04 36.3\hspace{0.2cm}  $-$33 59 18.8 &     & VLT$^{EJ}$ &3525.825 &     &  non measurable\\
\hline
IC 4637  = HD 154072     & 9 & & \multicolumn{3}{c}{round rim + shell + structures; Possible visual binary}\\
17 05 10.5\hspace{0.2cm}   $-$40 53 07.8 &  & VLT$^{EJ}$ & 3527.794   & 10 & $ 420\pm 335$\\
\hline
ESO588-14 = PN G008.2+06.8  &  11 & & \multicolumn{3}{c}{ bipolar; likely young - \bf emission lines} \\
17 38 57.1\hspace{0.2cm}  $-$18 17 35         &                    &  VLT &    6173.564           &  &  non measurable\\
\hline
Tc 1     = HD 161044       & 11.3 & &  \multicolumn{3}{c}{round with halo} \\
17 45 35.3\hspace{0.2cm} $-$46 05 23.7   &  &  VLT$^{EJ}$ & 3525.889 &  11 & $  -55\pm  330 $\\
                                          & &  VLT$^{EJ}$ & 3527.921 &  13 & $ -1845\pm 1645$\\
\hline
NGC 6629 = HD 169460    & 11.9 & &  \multicolumn{3}{c}{elliptical;  shell;  ISM distorted halo -  \bf emission lines}       \\
18 25 42.4\hspace{0.2cm} $-$23 12 10.2         &                    &  VLT &   6173.676/6174.626/6175.682  &  &  non measurable\\
\hline
IC 4776  = HD 173283    & 10.6 & &  \multicolumn{3}{c}{bipolar + knots - \bf emission lines}       \\
18 45 50.6\hspace{0.2cm} $-$33 20 32          &                    &  VLT &   6173.596/6176.637            &  &  non measurable\\
\hline
NGC 6826 = HD 186924       & 10.2 &     & \multicolumn{3}{c}{elliptical; low ionization structures + shell + halo}  \\
 19 44 48.2\hspace{0.2cm} $+$50 31 30.3                          &      & WHT & 4011.383  &  1.6 &  $  -660 \pm    390$\\
                           &      &     & 4012.375  & 11.2 & $ 1325  \pm     1575 $\\
\hline
NGC 7009 = HD 200516    & 12.5  &  &  \multicolumn{3}{c}{elliptical;  shell; jets;  ansae;  knotty halo - \bf emission lines }       \\
21 04 10.9\hspace{0.2cm} $-$11 21 48.3          &                    &  VLT &  6173.709/6174.735/6175.722             &  &  non measurable\\
\hline
NGC 7293 = Helix nebula   &  13.2  &  &  \multicolumn{3}{c}{pole-on inner disc; cometary knots; multiple bipolar outflows}  \\
22 29 38.6\hspace{0.2cm} $-$20 50 13.6                 &        & VLT$^E$ & 3527.886  & 6.5 & $ 1730 \pm   3150$\\
\hline
\multicolumn{6}{l}{\bf $^E$ESO Archive data }\\
\multicolumn{6}{l}{\bf $^J$AlsoJordan et al. (2012).}\\
\end{tabular}
\end{table*}

\section{Discussion}
Despite we have no positive detections of a magnetic field in any of
the observed PNe, this is significant result as targets span a large
range of nebular and stellar parameters (Table~\ref{tab:nebulinfo}).
In particular, a variety of morphological structures seen at different
inclination angles are included: from the marked bipolar shape of
HD~44179 (the Red Rectangle), the mild bipolar morphology of NGC~4361,
the ring/disc like inner nebula of NGC~7293 with outer multipolar
lobes, the elongated geometry of e.g. NGC~1360, to the almost
perfectly spherical shape of nebulae like Tc~1. In addition, highly
collimated structures which might be related to magnetic shaping are
also present, such as the highly inclined jets of NGC~1360 and the
pole-on ones of NGC~2392, or the symmetric pair of low-ionization
knots of NGC~6826. The only morphological class which is not well
represented (except for the case of the pre-PN HD~44179) is that of
classical ``butterfly'' nebulae, with a narrow waist from which high
velocity bipolar lobes depart (see e.g. Corradi \& Schwarz 1995).
While their extreme collimation put them among the most promising
targets where to look for magnetic fields \cite{Sabin07}, they are
beyond the reach of the method adopted in this paper. One reason is
that that their central stars are often intrisically faint, because
they are relatively massive and therefore have a fast post-AGB
evolution toward low luminosities (Corradi \& Schwarz 1995). In
addition, they have dense equatorial torii of gas/dust whose
emission/absorption often prevents the observations of the central
stars.

Some binary central stars, which can naturally provide additional
mechanism to produce magnetic fields (such as as angular momentum
transfer to one of the two stellar components) are also
included. NGC~6026 is a close binary with a period of 0.53~day
(Hillwig \citealt{h10}), composed of two degenerate compact objects
which experienced a common-envelope phase before ejecting the observed
nebula and jets. The period of NGC~1514 is not know, but the presence
of an A-type star at its centre, too cool to produce the ionization of
the nebula, indicates that it is also a binary system with likely a
much longer period as no large radial velocity shifts has been
detected so far \citep{Demarco04}. As the A-type
companion star dominates the spectrum in the visual range, our study
effectively looks for a magnetic field in the companion rather than in
the star that ejected the PN. This is still interesting, as the
possibility of an induced magnetic field in the companion stars, or
even of a common magnetosphere involving the two stars
(e.g. $\beta$~Lyrae, Leone et al. \citealt{Leone2003}),
exists.  Note also the longer period of HD~44179 (322 days), and the
possible very long period binary nature of IC~4637 \citep{Demarco04}.

Target nebulae are also peaked up at different evolutionary stages,
from the pre-PN nature of HD~44179, to the very evolved nebulae of
LSS~1362 and Abell~36. In this respect, it should be noted that the
nebula observed around PHL~932 might not be a PN, but just ambient
medium ionized by a hot sdB star, i.e. a Str\"omgren sphere (Frew et
al. \citealt{Frew10}).

Thus, in spite that our search for magnetic fields has addressed a
variety of morphological structures and central stars
parameters/duplicity, which have often been related to magnetic
fields, we have still no evidence for magnetic fields of the order of
a kG or somewhat less in PN central stars.
Following Kolenbergen \& Bagnulo \cite{Kolenbergen2009}, if the central stars of the PNe
here reported are all characterised by the same dipolar strength, with magnetic axis randomly oriented with
respect to the line of sight, then there is the 95\% probability that their dipolar strength is $< 800$ G,
while using Jordan et al. \cite{Jordan12} results
the upper limit is 1100 G. A forthcoming paper where a rigorous statistical analysis of
PNe observed so far from different groups will be presented (Asensio Ramos et al. 2014).  Future efforts will be
directed to detect these fields in the nebular gas and dust, which
will allow us to explore different types of targets and nebular
parameters.

\begin{acknowledgements}
AAR, MJMG and RMS acknowledge financial support by the Spanish
Ministry of Economy and Competitiveness (MINECO) through project
AYA2010-18029, and RLMC through proyect AYA2012-35330. AAR and RMS
acknowledge financial support by the MINECO through the project
Consolider-Ingenio 2010 CSD2009-00038. This paper is based on
observations made with the WHT operated on the island of La Palma by
the ING in the Spanish Observatorio del Roque de los Muchachos of the
Instituto de Astrof\' isica de Canarias. It is also based on observations made
with ESO Telescopes at the La Silla Paranal Observatory under programme ID 089.D-0429(A).
We thanks the referee, Dr. Stefano Bagnulo, for suggestions and comments.
\end{acknowledgements}

\end{document}